%% file: main.tex
\documentclass[11pt]{article}
\usepackage{preamble}

\title{Modeling Engagement with Brand and Organizational TikTok Videos Using Machine-Assisted \\ Theory-Ensemble Annotation}

\author{
Sander Paekivi\textsuperscript{1} \hspace{0.7mm} 
Andres Karjus\textsuperscript{1,2,3} \\
 \textsuperscript{1}Tallinn University \hspace{0.7mm} \textsuperscript{2}University of Tartu \hspace{0.7mm}  \textsuperscript{3}Estonian Business School
}
\date{\vspace{-0.7cm}}

\begin{document}

\maketitle

\begin{abstract}
Short-form video is difficult to study at scale because meaning emerges through audiovisual elements, language, and participatory, algorithmic and trend-based platform dynamics. Manual annotation of these layers is laborious at scale and difficult to standardize. We demonstrate how multimodal large language models (LLMs) can help address this bottleneck by annotating a set of 77 theory-driven structural variables derived from narratology, rhetoric, communication, and semiotics. We use this to explore content and estimate engagement with modest but consistent gains over account-size and video-age baselines in a corpus of about 10,000 TikTok videos of brand and organizational accounts from Estonia (covering a substantial share of the small country ecosystem). Human validation shows a reliability gradient: perceptual and communicative variables can be coded fairly reliably, while deeper semiotic and archetypal constructs are more difficult for both humans and machines. This approach of computational operationalization of long-standing interpretive theories can support several aims: exploratory cultural analytics of variation in short-form video culture, predictive modeling of platform dynamics, engagement, and audience feedback; and diagnostics for content creators to support choosing between structural and narrative strategies. Most annotated variables were not associated with platform success, as expected; the value of LLMs in this setting lies in making it feasible to assess large batteries of theoretically motivated variables, so that the subset carrying signal can be identified and translated into creator-facing guidance for a given niche.
\end{abstract}


\section{Introduction}

Online platforms hosting short-form videos have become a major site of cultural production, public communication, and commercial persuasion. Short videos on platforms such as TikTok, YouTube or Instagram are platform-native audiovisual content whose meanings emerge through the combination of image, speech, text overlays, music, editing, trend dynamics, and algorithmically shaped circulation \parencite{Lin2023,Grzenkowicz2025}. Such content is brief, highly multimodal, and often built around conventions of direct address, trend participation, stylized authenticity, and rapid genre switching \parencite{Jia2024,Christiansen2025}. For brand and organizational accounts, this matters because audiovisual storytelling can shape not only attention but also downstream organizational perception, and brand equity and experience \parencite{Teraiya2023}.

Existing large-scale research on virality and engagement has investigated this, but it has often privileged predictors that are comparatively easy to measure at scale: social influence, network structure, timing, account size, or general multimodal feature sets \parencite{Hidayati2017,Shang2022,Sangiorgio2025}. Work on cultural markets and online diffusion shows why these dynamics are difficult to absorb into static content features: social influence, random copying, and early network spread can make popularity path-dependent \parencite{salganik_experimental_2006,bentley_regular_2007,weng_virality_2013,goel_structural_2016}. Recent TikTok and short-video research has linked engagement to sentiment and second-person address in news videos, brand perception to user engagement and sentiment trends in platform-specific brand content, audiovisual features such as colloquial expression, cadence, colorfulness, and visual prominence in TikTok advertising, and multimodal information features in short sales videos \parencite{cheng2024like,martiochoa2025airbnb,zhang2025audiovisual,xiao2025multimodal}. Engineering-oriented studies show that multimodal LLMs can improve short-video engagement prediction, especially when the audio component is retained \parencite{sun2025engagement}. This is valuable for forecasting and content planning, but does not fully answer interpretive questions about why a video engages its audience, what kinds of symbolic work it performs, or how it positions viewers. In parallel, qualitative studies have shown the value of close analysis for health communication, misinformation, identity, and visual storytelling, but usually rely on comparatively small samples, given the labor-intensive nature of multimodal annotation \parencite{Li2021,Lookingbill2022,Southerton2022,Tudehope2024,Woolard2024,Christiansen2025}.

LLM pipelines have already been shown to support short-form video understanding at scale, including hook analysis in ads, mental-health discourse, influencer-value extraction, and domain-specific TikTok thematic analysis \parencite{yang2026mllmvadstory,zhang2026decoding,zha2026interpreting,starovolovsky-shitrit2025value,ghosh2025humanai}. Related computational social-science benchmarks show competitive performance on structured social-media annotation, including contextual interpretation, but also sensitivity to model, task, prompt, and validation design \parencite{tornberg2025large,stromergalley2026efficacy,lin2025navigating}. The methodological opening is a workflow for scaling qualitative coding--including thematic, narrative, content, linguistic-feature, discourse, and social analysis--through explicit codebooks, validation samples, replicable prompts, and bias checks, so that qualitative distinctions can become structured variables without losing interpretability \parencite{dunivin2025scaling,cui2025bias,Thapa2025,Tai2024,Jenner2025,Wen2025,Ziems2024,Karjus2025}.

\begin{figure}[t]
\centering
\includegraphics[width=\textwidth]{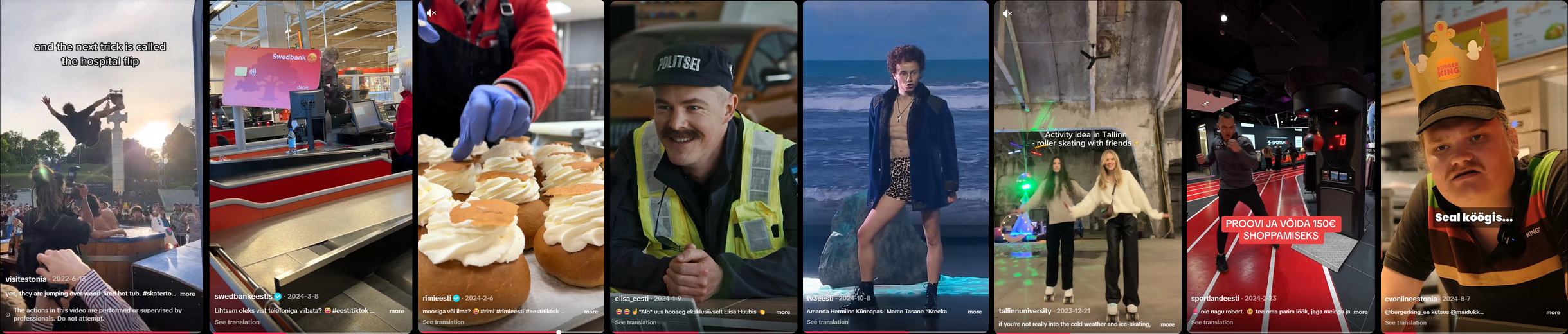}
\caption{Examples of TikTok videos in the corpus, from tourism and retail promotions to food, public-facing organizational content, music, activity prompts, and service advertising.}
\label{fig:examplevideos}
\end{figure}

The contribution here is a theory-saturated, forced-choice coding design, applied to the analysis of TikTok videos posted by a curated sample of brand, organizational, and enterprise-representing accounts (see Figure~\ref{fig:examplevideos} for examples) from Estonia, a small country of 1.3 million inhabitants in North-Eastern Europe. The linguistically and culturally bounded sample makes it possible to examine one delimited platform repertoire rather than rely on isolated viral cases or unrelated videos sampled without context. The design remains human-verifiable, reproducible, and auditable, while yielding categories that can support cultural-analytic mapping and engagement modeling tied to recognizable production features. This study operationalizes a wide set of theoretical lenses from media studies, narratology, rhetoric, semiotics, and multimodal analysis into a structured annotation scheme. 

The broad codebook functions as a screening device: it makes many candidate descriptors observable, then lets the models show which ones carry signal for this corpus and which mainly remain descriptive. It focuses on durable dimensions of narrative structure over short-lived platform trends, operationalizing rhetorical address, semiotic organization, and viewer positioning that have long been used to study creative works across art, literature, film, and other creative media. The operationalization follows a machine-assisted quantitizing (also known as converting or transforming) design logic in which qualitative judgments are rendered as quantifiable (in practice mostly categorical) variables while retaining their theoretical provenance and the possibility of expert review \parencite{sandelowski_quantitizing_2009,Karjus2025}. Related computational social-science work has likewise treated narrative structure as an object that can be operationalized and modeled computationally \parencite{yilmaz2022narrative}.

The study addresses three research questions:

\begin{enumerate}
\item How can multimodal, narrative, rhetorical, and semiotic theories be operationalized into a structured LLM annotation scheme for TikTok videos?
\item Which annotated features and broader structural patterns are most informative for predicting engagement and exploratory audience-response outcomes in Estonian brand and organizational TikTok videos after controlling for account size and video age?
\item What methodological gains and risks follow from using LLMs to annotate large batteries of short-form audiovisual variables?
\end{enumerate}

\section{Data and Methods}

\subsection{Corpus Construction and Sampling}

The corpus is purposive rather than statistically representative. The set of accounts to mine for video content was identified and curated through a combination of Estonian keyword and hashtag searches (e.g., \textit{\#estonia}, \textit{\#eesti}), manual scouting of brand accounts, and recommendation trails within TikTok. The aim was to assemble a large corpus of Estonian brand or organizational accounts with substantial video archives rather than to estimate population parameters for broader TikTok content. The merged dataset contains 9,654 videos from 30 accounts. Videos in the corpus were posted between 7 July 2020 and 10 January 2026, although most date from 2023--2025. Account archives vary substantially in size, and engagement is likewise highly skewed, with many videos attracting modest interaction and a smaller number reaching very high totals, as is typical for social media content. Account-level follower counts were captured at scraping time and repeated across all videos from the same account, so they are treated as account-level snapshot controls rather than time-varying exposure measures.

\begin{table}[htbp]
\centering
\caption{Overview of the TikTok corpus}
\label{tab:corpus}
\begin{tabularx}{\textwidth}{>{\raggedright\arraybackslash}p{0.52\textwidth} >{\raggedleft\arraybackslash}X}
\toprule
Measure & Value \\
\midrule
Accounts & 30 \\
Videos per account & min=18, median=266, max=894 \\
Followers per account & min=121, median=9,950, max=44,700 \\
\midrule
Videos & 9,654 \\
Data mining period & 2020-07-07 to 2026-01-10 \\
Share of videos from 2023--2025 & 83\% \\
Likes per video & min=0, median=253, max=316,300 \\
\midrule
Coded content variables & 77, in 11 blocks \\
Variables used in predictive analyses & 75 content, 2 control \\
Successfully LLM-annotated videos & 9,649 \\
\midrule
Comments per video & min=0, median=3, max=15,900 \\
Videos analyzed for comments (where present) & 5,804 \\
Top-level comment chains retained & 85,484 \\
Replies retained within comment chains & 24,114 \\
\bottomrule
\end{tabularx}
\end{table}

\subsection{Scraping and Metadata}

The corpus was assembled with a Python web-scraping workflow using Selenium WebDriver and Google Chrome in headed mode. For each account, the scraper collected account-level metadata, enumerated video URLs, visited videos one by one, extracted engagement counts and textual metadata, opened comment threads, and downloaded the corresponding local video file. 
The comment scraper targeted rich discussion data rather than a single headline metric. The corpus contains a maximum observed value of 219 top-level comment chains for a single video and up to 20 replies for a single comment chain. In practice, the median video contains only two top-level comment chains, which reflects the skewed engagement not unexpected of social media data.
Scraping took place in multiple waves between 1 July 2025 and 12 January 2026. We control for video age in statistical modeling by deriving the difference between upload date and scrape date. This is necessary because engagement counts are cumulative snapshots rather than same-age observations.

\subsection{LLM Annotation Design}

The theoretical coding scheme is implemented for Gemini 2.5 Pro to return a single JSON object and to avoid free-form prose outside the schema. Because model comparison was outside the scope of this study, we used a single video-capable frontier model. Each video is analyzed into two kinds of output: a generated description of the video and spoken-audio transcript, and a structured annotation across 11 theory blocks of 77 categorical or binary variables (\enquote{Not Applicable} is used as a logical escape value for dependent fields), and an additional 20 diagnostic explanatory text fields. The blocks cover general video content characteristics, Grzenkowicz and Wildfeuer's multimodal annotation framework for short video, rhetorical and audience address, Chandler-style semiotic analysis, Hall's encoding/decoding model, uses and gratifications, Burke's dramatistic pentad, Aristotelian rhetorical appeals (ethos, pathos, logos), a Propp-/Campbell-inspired arc measure, Labovian narrative structure, and an abstracted Polti conflict type \parencite{Hall2009,Livingstone1993,blumler1974,Sundar2013,Chandler2025,labov1972language,Burke1978,Aristotle1985,Lessard2022,Brusentsev2012,Campbell2008,Cowen2020,Han2022,hevner1936experimental,zentner2008emotions,Grzenkowicz2025}. Of the 77 variables, two binary variables (presence of music, text) are excluded in the predictive models as they are collinear with their more detailed categorical variables.
The theoretical ambition is deliberately broad. The prompt was designed to translate established concepts into finite choice sets that can later be modeled quantitatively, instead of asking the model for unconstrained interpretation.
The design does not assume that every short video instantiates every theoretical system. It instead follows an ensemble logic: multiple well-established frameworks are used to probe partially overlapping dimensions of a video's communicative core, after which empirical analysis can show which lenses are robust, which are redundant, and which add little signal in this domain. This is especially relevant for short-form video, where promotional, narrative, semiotic, and rhetorical functions often coexist in compressed form.

Table~\ref{tab:schema} summarizes the distribution of variables across sections.

\begin{table}[htbp]
\centering
\caption{Theory blocks and variables.}
\label{tab:schema}
\begin{tabularx}{\textwidth}{>{\raggedright\arraybackslash}X >{\raggedleft\arraybackslash}p{2.2cm}}
\toprule
Theory block & Schema fields \\
\midrule
General video characteristics (e.g. communicative intent, overall mood) & 21 \\
Grzenkowicz \& Wildfeuer multimodal (e.g. camera framing, audio type) & 25 \\
Rhetorical and audience analysis (e.g. address directness, direct gaze) & 6 \\
Chandler semiotic analysis (e.g. dominant sign mode, anchorage status, irony) & 16 \\
Hall encoding/decoding analysis (e.g. dominant ideology) & 3 \\
Uses and gratifications analysis (primary gratification, description) & 2 \\
Burke dramatistic analysis (e.g. act, scene, dominant element) & 7 \\
Aristotle rhetorical appeals (e.g. ethos, pathos, dominant rhetorical appeal) & 5 \\
Propp/Campbell narrative arc abstraction (arc shape, justification) & 2 \\
Labov narrative analysis (e.g. abstract hook strategy, coda call-to-action type) & 8 \\
Polti dramatic situations abstraction (core conflict, conflict justification) & 2 \\
Video metadata at scrape time (video age, account follower count) & 2 \\
\midrule
97 variables total: 75 predictors, 2 controls, and 20 audit variables & \\
\bottomrule
\end{tabularx}
\end{table}

\subsection{Annotation Validation}

Before the LLM annotations can be used to draw conclusions about engagement, their reliability must be evaluated by asking whether human annotators can agree on each variable's application to a given video and whether the LLM annotations match the human consensus. If humans cannot agree on a variable, no stable ground truth exists against which to evaluate the LLM results, and findings relating to the variable should be treated with caution. 

A subset of 100 videos was sampled from the full corpus for human validation by three paid student annotators. They completed the assignment partially rather than in full, coding 40, 73, and 58 videos, respectively, from the same shared assignment pool, with overlapping coverage across videos. The resulting overlap structure yields 57 videos with two or more independent human annotations and 1,133 LLM-to-human comparison pairs across the 77 variables. Agreement metrics are computed pairwise within each video on whichever annotators are present, rather than assuming a fixed pair across all items. 

Annotation reliability was evaluated in two steps. First, human agreement was summarized with raw percent agreement, Krippendorff's $\alpha$, and Fleiss' $\kappa$, with the latter two included as they adjust for chance agreement under skewed category distributions. Where a human majority label was available, the LLM annotations were evaluated against this benchmark using accuracy and macro-averaged F1. Accuracy gives the direct match rate, while macro-F1 gives equal weight to all categories and therefore penalizes models that appear accurate mainly by defaulting to frequent labels. These metrics were used as a compact diagnostic assessment of annotation quality rather than as a definitive validation of the full theoretical codebook.

Table~\ref{tab:validation} summarizes the validation results by theory block. The pattern reveals a reliability gradient that tracks the abstraction level of the underlying constructs.

\begin{table}[ht]
\centering
\caption{Human validation of LLM annotations by theory block. Inter-annotator agreement is reported as Krippendorff's $\alpha$ and Fleiss's $\kappa$ across 57 multiply-coded videos. LLM performance is evaluated against human majority vote using accuracy, alongside macro-averaged precision, recall, and F1 score.\textsuperscript{a}}
\label{tab:validation}
\small
\begin{tabular}{l rrr rrrrr}
\toprule
 & \multicolumn{3}{c}{Human agreement} & \multicolumn{4}{c}{LLM vs.\ humans} & \\
\cmidrule(lr){2-4} \cmidrule(lr){5-8}
Theory block & $\alpha$ & $\kappa$ & \%Agree & Acc & macP & macR & macF1 & $N$ \\
\midrule
Audiovisual: audio        &  0.58 &  0.67 & 81\% & 79\% & 0.55 & 0.55 & 0.54 & 136 \\
General: presenter        &  0.45 &  0.38 & 73\% & 80\% & 0.54 & 0.57 & 0.54 & 128 \\
Aristotle rhetoric        &  0.59 &  -0.34 & 86\% & 77\% & 0.60 & 0.65 & 0.59 &  60 \\
Audiovisual: editing      &  0.45 &  0.32 & 63\% & 72\% & 0.59 & 0.59 & 0.52 & 117 \\
General: content          &  0.57 &  0.43 & 73\% & 61\% & 0.44 & 0.45 & 0.42 & 119 \\
Audiovisual: visual       &  0.63 &  0.44 & 78\% & 79\% & 0.41 & 0.45 & 0.40 & 112 \\
Audiovisual: participants &  0.34 &  -0.18 & 58\% & 77\% & 0.42 & 0.40 & 0.36 &  48 \\
Narrative (Labov)         &  0.40 &  0.42 & 58\% & 63\% & 0.35 & 0.37 & 0.34 & 127 \\
Rhetorical address        &  0.59 &  0.53 & 65\% & 46\% & 0.33 & 0.41 & 0.33 &  69 \\
Hall / B\&K / Polti       &  0.12 &  -0.02 & 46\% & 50\% & 0.34 & 0.34 & 0.28 &  38 \\
Chandler semiotics        & -0.09 & -0.10 & 64\% & 54\% & 0.30 & 0.41 & 0.24 &  95 \\
Burke dramatism           &  0.07 &  0.23 & 57\% & 57\% & 0.31 & 0.21 & 0.20 &  84 \\
\midrule
Overall                   &       &       &      & 68\% & 0.41 & 0.43 & 0.39 & 1133 \\
\bottomrule
\end{tabular}

\vspace{4pt}
\raggedright
\end{table}

The audiovisual and general content blocks showed good performance on both dimensions. The audio block achieved the highest inter-annotator agreement ($\alpha = 0.58$, $\kappa = 0.67$, 81\% agreement) alongside 79\% LLM accuracy and a macro-F1 of 0.54. The presenter block (80\% accuracy, macF1 = 0.54) and visual block ($\alpha = 0.63$, 79\% accuracy, macF1 = 0.40) performed similarly. Within these blocks, several individual variables reached ceiling reliability: visual source type, primary audio layer, camera angle, speed alteration, and visual filter presence all achieved $\alpha = 1.0$ and LLM accuracy above 87\%. These variables capture directly observable distinctions, such as audio type, camera orientation, and visual filtering.

The Aristotle rhetoric block presents a somewhat mixed case: high percent agreement (86\%) and relatively strong LLM performance (77\% accuracy, macF1 = 0.59), but a divergence between $\alpha$ (0.59) and $\kappa$ (-0.34). This divergence is likely an artifact of low variance rather than genuine disagreement: two of the three binary variables (ethos and logos presence) had perfect agreement, but with all annotators selecting the same value. This inflates $\alpha$ but leaves $\kappa$ undefined or negative. The editing block ($\alpha = 0.45$, 72\% accuracy, macF1 = 0.52) and Labov narrative block ($\alpha = 0.40$, 63\% accuracy, macF1 = 0.34) also fall into this moderate range. These involve interpretive judgment, but are still grounded in relatively concrete cues, such as editing pace, the presence of a call to action, or a recognizable opening hook.

The deeper interpretive frameworks showed poor reliability on both dimensions, and the gap between accuracy and macF1 is particularly revealing. Burke's dramatistic pentad achieved 57\% accuracy but only a macF1 of 0.20, indicating that the LLM is defaulting to common elements (Act, Agent) while failing entirely on more nuanced ones (Agency, Purpose). The inter-annotator agreement ($\alpha = 0.07$) supports this, demonstrating that this is not merely an LLM failure; human annotators were also unable to agree on which pentad element dominates a given video, producing agreement barely above chance. The Chandler semiotics block ($\alpha = -0.09$, $\kappa = -0.10$, macF1 = 0.24) is the clearest case of systematic disagreement, where negative $\alpha$ values indicate that annotators interpreted semiotic categories in opposing ways more often than random assignment would predict. The Hall, Blumler and Katz, and Polti block ($\alpha = 0.12$, 50\% accuracy, macF1 = 0.28) is similarly unreliable, with Polti's core conflict variable achieving only 12.5\% accuracy.

Beyond variable-level accuracy, the LLM exhibited three consistent patterns of disagreement with human coders. First, it systematically over-detected dramatic conflict: in 86\% of Polti disagreements, the LLM assigned a specific conflict type (e.g., "Quest/Daring Enterprise") to videos that humans coded as "No Overt Conflict". Second, it over-detected semiotic complexity, finding opposition, anchorage, and symbolic sign modes where humans perceived straightforward iconic representation. Finally, it defaulted to intimacy in audience address, coding creator-viewer relationships as more direct, familiar, and peer-like than human annotators perceived.

Annotators were non-specialist undergraduate students who received the general task description and the question text as their only guidance. Low agreement on theory-heavy variables (Burke, Peirce, Hall) may therefore partly reflect annotator inexperience rather than construct invalidity. However, the present validation cannot distinguish between these explanations, and future work should employ domain-expert annotators to establish reliability baselines for these constructs. Accordingly, findings from audiovisual and communicative packaging variables are treated as better supported in the validation subset, while findings from deeper interpretive variables are retained as exploratory indicators. Their predictive value can still be informative, but specific category interpretations from these blocks require caution because the validation exercise cannot confirm that the theoretical labels are being assigned faithfully.

\subsection{Quantitative Analysis Strategy}

The engagement outcomes are modeled as smoothed log-transformed counts (\texttt{log1p}) for likes, comments, shares, and saves/bookmarks, as are the two control variables of account follower count and video age. Two binary indicators, "music present" and "text overlay present", are not used in modeling because they are exact logical duplicates of levels in the related categorical variables.

The inferential design prioritizes grouped out-of-account prediction. In the first stage, grouped CatBoost models are estimated because they handle many categorical predictors naturally and do not require arbitrary reference categories. Account-level grouping is enforced in every train/test split so that videos from the same account never appear in both training and test partitions. Five grouped splits are used to estimate the mean and standard deviation of held-out performance. For each outcome, a controls-only model is compared with a full model that also includes the theory-coded content variables. This makes it possible to estimate the incremental predictive contribution of the annotations beyond baseline visibility controls: account follower count and video age.

In the second stage, up to ten of the most important content predictors from the first-stage CatBoost models are passed to a smaller linear mixed-effects model, alongside controls, with effect (sum) coding and an account random intercept. Effect coding means that category coefficients are deviations from the grand mean of each predictor rather than comparisons with one arbitrary baseline.

For likes and comments, an additional design-choice summary is derived from the same grouped account-split CatBoost procedure. In each of five grouped splits, CatBoost is fit on the training accounts only. On the held-out accounts, each predictor is overwritten in turn with each of its common levels while all remaining features, including follower count and video age, are left at their observed values. Common levels are defined as levels with at least 80 videos, capped at eight levels per variable. Each variable is then summarized by two split-averaged quantities: the contrast between its highest-performing common level and the mean prediction across its remaining levels, and the effective account coverage of that best level, expressed as a percentage of accounts. These summaries identify variables whose best-performing common level combines a larger modeled contrast with either broad or narrow distribution across accounts.

This design aims to separate exploratory feature discovery from directional effect estimation, preserves grouped out-of-account evaluation, and yields more interpretable results than a purely black-box forecasting exercise. At the same time, the analysis is associational rather than causal. Neither the purposive account sample nor the observational platform data justify strong claims about what causes engagement.

\subsection{Exploratory Mapping of Videos and Comments}

The predictive models are supplemented by two descriptive structure analyses on the same harmonized annotation table. The first represents each video directly by its coded variables and defines pairwise video dissimilarity by Hamming distance over those fields. A two-dimensional corpus map is then estimated once with a Hamming-based t-SNE layout and reused across all nine panels, which keeps the similarity structure at the level of the original variables rather than expanding higher-cardinality variables into larger one-hot blocks.

TikTok comment sections were available in the scrape as nested thread data attached to each video, including top-level comments and any visible replies. For comment-side analysis, each visible section was reconstructed in thread order, preserving comment--reply grouping while linearizing the text into blocks separated by empty lines. Only videos with at least one scraped top-level comment and at least 20 visible characters of comment text were retained, yielding a codable subset of 5,804 videos. Each reconstructed section was then submitted as a whole to GPT-5.4 via the API (temperature at 0, reasoning parameter at "none") and a constrained structured-output schema. The model assigned one dominant or summarizing comment-activity label per section: supportive, critical, information-seeking, neutral, or unclear. Supportive captured positive or enthusiastically participatory commenting, critical captured negative, sarcastic, or complaint-oriented commenting, and information-seeking captured requests for details, clarification, price, availability, or instructions. Neutral was reserved for matter-of-fact sections without clear affective or informational direction, and unclear for highly fragmentary or ambiguous sections. As a post-annotation check on this second LLM-coded layer, we manually reviewed an account-stratified sample of 20 comment sections and found the assigned dominant activity labels acceptable for exploratory analysis. These section-level labels were merged back into the video-level dataset and used for descriptive mapping and grouped account-holdout predictive modeling.

\section{Results}

\subsection{Exploratory Structure of the Annotated Space}

\begin{figure}[htbp]
\centering
\includegraphics[width=\textwidth]{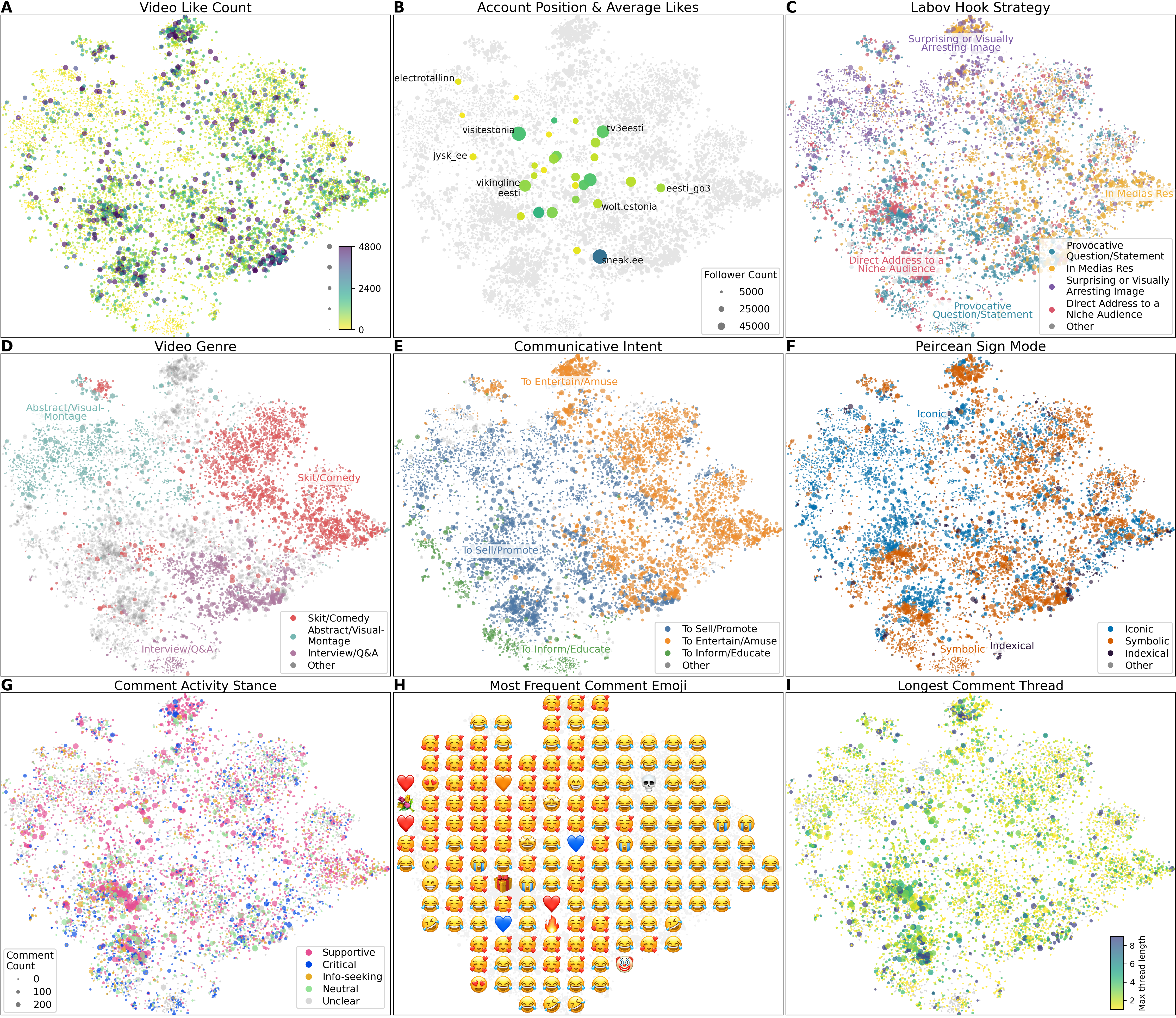}
\caption{Shared configuration space of 9,649 videos from 30 purposively sampled Estonian brand and organizational TikTok accounts. Each point is a video positioned by a fixed t-SNE layout over pairwise Hamming distances across all coded variables; size tracks like count in A--F and comment count in G--I. A: video likes. B: account centroids, sized by follower count and colored by mean likes on A's scale. C--F: Labovian hook strategy, video format, communicative intent, and Peircean sign mode. G--I: dominant comment stance, most common emoji, and maximal reply-thread length. Likes, comments, and thread length use clipped linear scaling at the 2nd and 98th percentiles, compressing tails without removing observations. The map reveals a structured but continuous repertoire: entertainment- and promotion-oriented configurations occupy partly distinct regions, and audience responses vary systematically across the same design space.}
\label{fig:corpusmap}
\end{figure}

Figure~\ref{fig:corpusmap} maps all 9,649 analyzable videos into a common configuration space derived from pairwise Hamming dissimilarities over the full annotation set. Because position is determined by all coded variables at once, the separation of video format and communicative intent in \ref{fig:corpusmap}D--E indicates that these practical schema variables align with the broader rhetorical, narrative, and semiotic structure of the corpus. The most common video formats are skit/comedy (27.9\%), abstract or visual montage (19.2\%), and interview or Q\&A (13.0\%). The dominant communicative intents are selling or promotion (48.3\%), entertainment or amusement (34.6\%), and information or education (8.4\%). These categories also organize the map: promotional videos are more concentrated in the left and lower-left regions, entertaining videos dominate much of the right half, and informative or educational videos form a smaller lower-left band. Genre sharpens the same structure, with comedy skits filling much of the entertainment-heavy right side, abstract visual montages appearing more often in promotional regions, and interview or Q\&A videos appearing more often in the lower band.

The audiovisual and narrative annotations point in the same direction. The primary audio layer is most often spoken dialogue or narration (55.8\%) or music (42.7\%), while rhetorical tone is most often humorous (46.3\%), persuasive or sales-oriented (22.7\%), or inspirational or motivational (13.9\%). The Polti core-conflict variable captures the video's dominant narrative tension, from explicit conflict to looser problem- or goal-oriented framing; its most common labels are quest or daring enterprise (32.9\%) and lack of overt conflict (32.2\%). Estonian brand and organizational TikTok in this corpus therefore looks less like a set of isolated account niches than like a shared repertoire organized around recurring promotional, entertaining, and explanatory poles. The account centroids in \ref{fig:corpusmap}B support that reading: although the 30 purposively sampled accounts occupy different neighborhoods, most remain within the shared middle of the repertoire rather than at its far fringes.

The theory-specific panels clarify the structure of these regions. The Labovian hook variable in \ref{fig:corpusmap}C captures the video's opening abstract or attention-securing move in the first 3--5 seconds, adapted from Labov's narrative model \parencite{labov1972language}. The most common hooks are provocative question or statement (29.3\%), \textit{in medias res} openings that begin in the middle of the action (25.7\%), and surprising or visually arresting images (22.3\%). Direct address to a niche audience is more concentrated in the promotional left-hand zones, whereas \textit{in medias res} and visually arresting hooks extend further into the entertainment-heavy right and upper regions. The Peircean panel in \ref{fig:corpusmap}F distinguishes whether the video's dominant signs work mainly through resemblance, direct trace, or learned convention \parencite{peirce1985logic,Chandler2025}. Iconic and symbolic sign use appears almost equally in this corpus (48.0\% and 46.6\%), while indexical signs are rare (5.4\%), suggesting that creators rely more on staged visual resemblance and culturally legible symbolic cues than on documentary trace. A video by the telecommunications company Elisa in the entertainment region exemplifies this: a deadpan exchange between a child and an increasingly frustrated man, playing as a short comedy skit that starts mid-action; it prompted audience comments such as \enquote{I'm dying of laughter}.

The comment-focused panels \ref{fig:corpusmap}G--I show that visible comment activity follows a similar design space. Of all videos, 60.2\% yielded codable comment sections, among which supportive are most common (47.9\%), followed by critical (18.1\%), neutral (17.9\%), and information-seeking (15.1\%). Entertaining videos produce supportive comment sections much more often than information-seeking ones (55.6\% versus 10.1\%), while informative videos and sell-or-promote clips attract questions (17.7\% and 18.2\%, respectively). Panel \ref{fig:corpusmap}H is dominated by laughter and affection markers, reinforcing the broadly playful or approving tone of much of the response ecology. Panel \ref{fig:corpusmap}I shows that longer reply chains do not only accumulate where likes are highest, but also emerge around videos that invite clarification, correction, or disagreement. A Sportland how-to clip, for example, draws practical questions such as \enquote{Can you do this even if you're not flexible?} and \enquote{Can you make a video on how to get into a bridge from standing?} By contrast, a Visit Estonia dance-promo clip attracted a predominantly critical thread with comments like \enquote{Please delete this} and \enquote{This dance ruined my FYP.} The map therefore situates comment activity within the same content-similarity space; the following sections model these associations statistically.

\FloatBarrier

\subsection{Predicting Content Performance Beyond Visibility Controls}

The central empirical question is whether theory-coded content variables improve prediction over the visibility controls, account follower count and video age. Table~\ref{tab:performance}A shows moderate improvement. In grouped held-out evaluation, the full CatBoost models reach \(R^2 = 0.267\) for likes, \(0.183\) for comment count, \(0.162\) for shares, and \(0.266\) for saves. Relative to controls-only models, the annotation variables add between \(0.056\) and \(0.078\) \(R^2\) across those engagement outcomes.
Likes and saves are the most predictable engagement outcomes in this corpus, whereas shares remain the hardest. The split-to-split standard deviations in Table~\ref{tab:performance}A show that performance varies across held-out account groupings even when the average increment from the content variables remains positive.

In the second-stage mixed models, only likes retained a clearly nonzero account random-intercept variance (\(\sigma^2 = 0.28\)), meaning that accounts still differed in baseline like levels after follower count, video age, and the strongest content variables were included. For comment count, shares, and saves, the account-level variance component was almost zero, indicating that this second-stage specification did not estimate additional stable between-account baseline differences beyond those predictors.

\begin{table}[htbp]
\centering
\caption{Held-out predictive CatBoost performance across five grouped account splits. Block A reports the main video engagement outcomes ($R^2$ and Root Mean Square Error), while Block B reports two comment-section activity outcomes on the subset of videos with sufficient comment sections (Area Under the Curve and Average Precision of the predicted class). In each cell, \(\pm\) gives the split-to-split standard deviation on the same metric scale as the reported mean.}
\label{tab:performance}
\begin{tabularx}{\textwidth}{l >{\raggedleft\arraybackslash}X >{\raggedleft\arraybackslash}X >{\raggedleft\arraybackslash}X >{\raggedleft\arraybackslash}X}
\toprule
\multicolumn{5}{l}{\textbf{A. Video engagement outcomes}} \\
Outcome & Full \(R^2\) & Ctrl \(R^2\) & \(\Delta\)content & RMSE \\
\midrule
Likes & \(0.27\pm0.13\) & \(0.19\pm0.10\) & 0.08 & \(1.25\pm0.06\) \\
Comment count & \(0.18\pm0.13\) & \(0.12\pm0.08\) & 0.07 & \(1.29\pm0.07\) \\
Shares & \(0.16\pm0.07\) & \(0.11\pm0.06\) & 0.06 & \(1.66\pm0.09\) \\
Saves & \(0.27\pm0.15\) & \(0.21\pm0.12\) & 0.06 & \(1.32\pm0.12\) \\
\addlinespace[0.4em]
\midrule
\multicolumn{5}{l}{\textbf{B. Comment-section activity outcomes}} \\
Outcome & Full AUC & Ctrl AUC & \(\Delta\)content & AP \\
\midrule
Supportive & \(0.65\pm0.05\) & \(0.62\pm0.05\) & 0.03 & \(0.61\pm0.09\) \\
Information-seeking & \(0.64\pm0.03\) & \(0.60\pm0.07\) & 0.05 & \(0.22\pm0.03\) \\
\bottomrule
\end{tabularx}
\end{table}

\subsection{Predicting Commenting Behavior}

The comment-section models in Table~\ref{tab:performance}B shift the target from comment volume to dominant comment-section activity (annotated using GPT-5.4). On the subset of 5,804 videos with codable comment sections, grouped CatBoost classifiers were estimated for whether the dominant visible activity in the section (analyzed as a whole) was supportive rather than all other coded options, and information-seeking rather than all other coded options (see also Figure \ref{fig:corpusmap}G). Predictive performance is weaker than for the engagement-count models, but still above the visibility-control baseline. Grouped AUC rises from \(0.62\) to \(0.65\) for supportive and from \(0.60\) to \(0.64\) for information-seeking comment sections, with average precision at \(0.61\) and \(0.22\). The models therefore suggest that the coded video features contain signal about both comment volume and the dominant visible pattern of comment-section activity.

Supportive comment sections are associated mostly with affective and audience-facing packaging variables, especially pathos-led rhetoric, creator-as-servant or entertainer address, call-to-action overlays, and topics such as travel or fashion/beauty. The clearest within-variable contrast is between pathos and logos: the split-averaged supportive-comment probability is about 10 percentage points higher for pathos, while several topic and address variables show additional four-to-six-point contrasts between their strongest and weakest levels. Information-seeking comment sections are more associated with selling or promoting intent, tutorial or how-to format, logos-led rhetoric, neutral or expert-style address, and more agency-centered framing. The absolute probability shifts are smaller, usually around 1-3 percentage points above the average choice mix, but remain consistent enough across grouped splits to mark a response pathway: entertaining, affective packaging is somewhat more likely to be followed by supportive comment-section activity, whereas practical and explanatory packaging is more likely to be followed by questions and requests for information.

\begin{figure}[htbp]
\centering
\includegraphics[width=\textwidth]{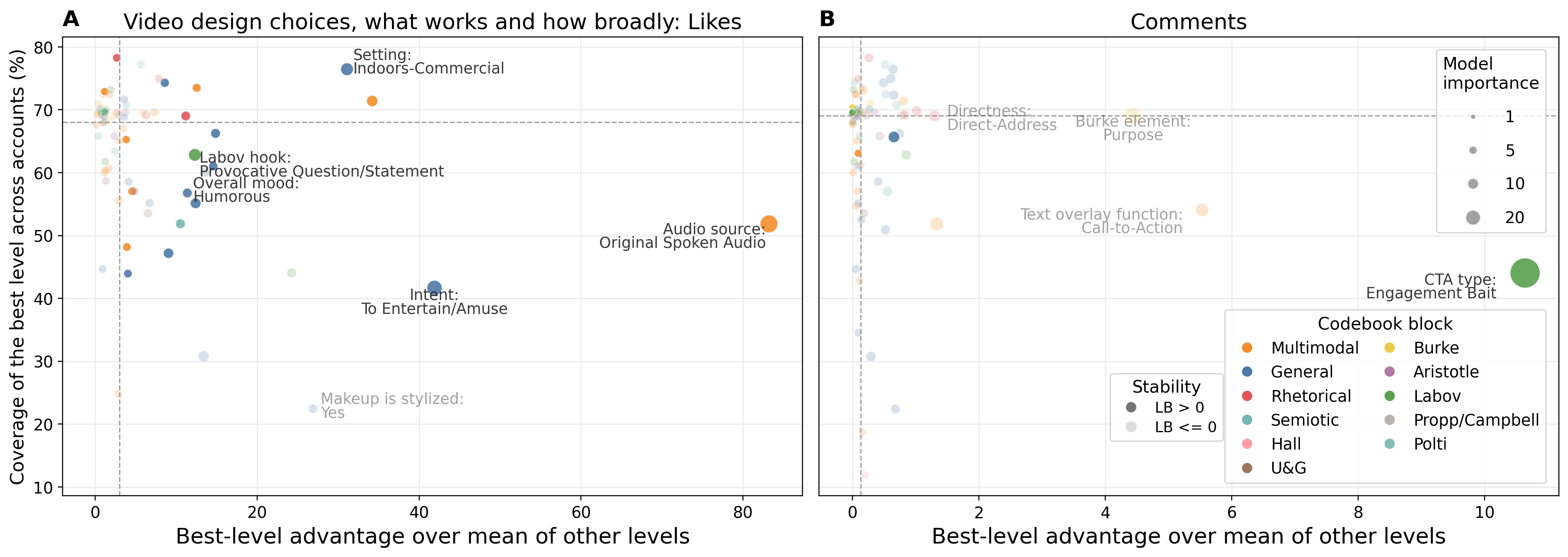}
\caption{Modeled design contrasts across breadth and stability for likes (A) and comment count (B). Each point is one predictor variable, colored by codebook block. The horizontal axis shows the split-averaged contrast between the highest-performing common level of that variable and the mean prediction for its remaining levels in grouped account-split CatBoost scenarios on held-out videos; the vertical axis shows how widely that best level is distributed across accounts. Point size shows relative variable importance. Transparency is binary: opaque points have an approximate split-based 95\% lower bound above zero, while transparent points do not. For likes, audio source, intent, and setting combine relatively large modeled contrasts with broad coverage, whereas stylized makeup is narrower and less stable. For comments, call-to-action (CTA) type has the largest modeled contrast, led by engagement bait.}
\label{fig:likescenarios}
\end{figure}

\subsection{Modeled Design Choices and Their Effects on Likes and Comment Counts}

Figure~\ref{fig:likescenarios} summarizes the full modeled variable set in the grouped out-of-account CatBoost framework. The horizontal axis compares the highest-performing common level of each variable with the mean prediction across its remaining levels, while the vertical axis shows how broadly that best level appears across accounts. Here, "design choice" is an analytic shorthand for observable coded alternatives in the videos, not a claim that creators consciously think in these categories. Because each scenario changes only one variable at a time while leaving all others at their observed values, the raw-count contrasts should be read as marginal model-implied differences rather than complete design prescriptions. For scale, the median modeled video receives 253 likes (IQR: 92--692) and 3 comments (IQR: 1--11), so shifts of a few dozen likes or several comments are modest in platform-wide terms but meaningful within this corpus of Estonian brand accounts.

In the likes panel (A), audio source has the largest modeled contrast: original spoken audio yields a split-averaged contrast of about 83.2 likes over the other audio-source levels (\(SD=17.7\)), with effective coverage of 51.8\% of accounts. Communicative intent follows at 41.9 likes (\(SD=15.5\)), led by entertaining or amusing intent. Setting is broader rather than sharper: indoors-commercial settings show a 31.1-like contrast (\(SD=10.3\)) and appear across 76.4\% of accounts. By contrast, stylized makeup shows a smaller and less stable niche pattern: 26.9 likes, 22.4\% account coverage, and high split variability (\(SD=25.7\)). The models indicate that likes are most strongly associated with broad communicative packaging choices such as audio source, intent, visual-filter state, and setting, while some stylistic signals are narrower or less stable.

Call-to-action type dominates both the modeled contrast and model importance in the comments panel (B). Engagement bait yields a split-averaged contrast of 10.64 comments over the other CTA levels (\(SD=5.54\)) with 44.0\% account coverage. Text-overlay function is next at 5.53 comments (\(SD=6.22\)), but it is shown with transparency because the estimate is less stable across grouped account splits. Burke's dominant element follows at 4.44 comments (\(SD=4.59\)) with broad coverage at 69.0\%, led by purposive framing. Many variables remain near the lower-left corner in one or both panels, indicating limited incremental leverage once follower count, video age, and stronger multimodal packaging variables are taken into account.

\subsection{Structural Similarity, Distinctiveness, and Relative Performance}

\begin{figure}[htbp]
\centering
\includegraphics[width=\textwidth]{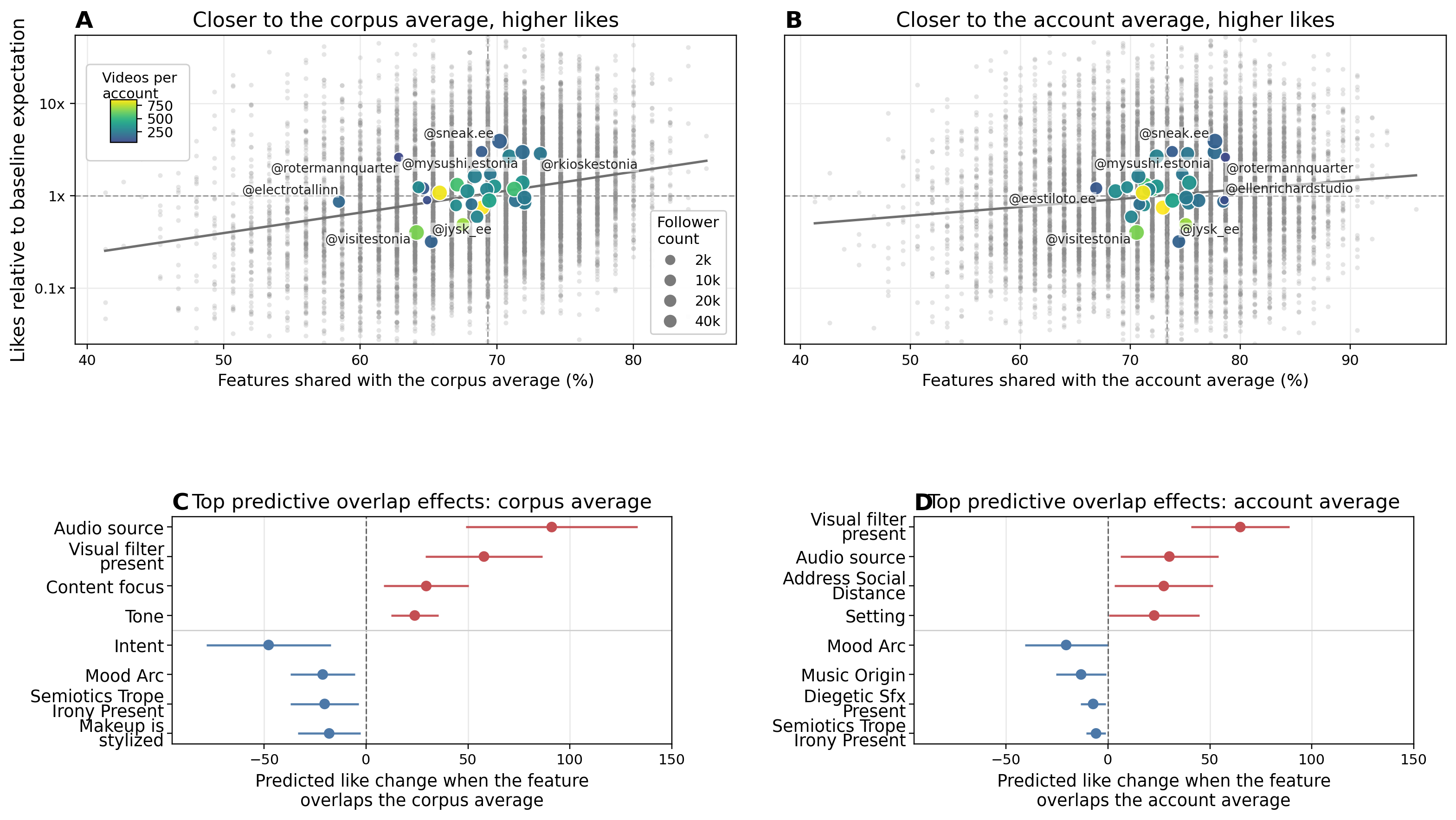}
\caption{Structural similarity, modal-profile agreement, and relative like performance. Panels A and B summarize each video's full coded profile as the percentage of the 75 modeled content features whose values agree with either the corpus-modal or account-modal profile. Because the features are categorical, the \enquote{modal profile} is the reference profile formed by taking the most common value of each feature, either across the full corpus or within each account. Small gray points are individual videos; larger colored points are account averages, sized by follower count and colored by the number of videos available for that account. The vertical axis rescales residuals from a controls-only model of \(\log_{10}(\mathrm{likes}+1)\) on follower count and video age so that \(1\times\) marks the baseline expectation, values above it indicate overperformance, and values below it indicate underperformance. 
Panels C and D decompose this aggregate result with grouped account-split CatBoost scenarios: for each variable, a binary modal-agreement indicator is toggled from 0 to 1 on held-out videos while all remaining features are left unchanged. The four strongest positive and negative stable variables are shown here. Points show the resulting average predicted change, and horizontal lines show approximate split-based 95\% intervals. Red points indicate that agreement with the modal value is favorable; blue points indicate that divergence is more favorable. Greater structural similarity is positively associated with relative like performance overall, but the lower panels show that this net pattern is carried mainly by a subset of high-leverage packaging variables rather than by uniform conformity across the entire codebook.}
\label{fig:distinctivenesslikes}
\end{figure}

To assess whether distinctiveness or originality, as captured by the coded variables, is rewarded, Figure~\ref{fig:distinctivenesslikes} summarizes each video's coded profile as a modal-profile agreement score. Panels A and B report the percentage of the 75 modeled content features whose values agree with either the corpus-modal profile or the account-specific modal profile. The reference profile is not a numeric average, but the most common coded value for each feature. The vertical axis reports like performance relative to a baseline expectation from follower count and video age. Panels C and D then decompose the same question at the variable level by asking which specific modal-value agreements raise or lower predicted likes in grouped account-split CatBoost scenarios.

The relationship is positive in both formulations. In a mixed-effects model with account random intercepts, similarity to the corpus-average profile is positively associated with baseline-adjusted likes (\(\beta = 0.123\), \(SE = 0.006\), likelihood-ratio \(\chi^2(1) = 433.91\), \(p < .001\)). Similarity to the account-average profile is also positive but weaker (\(\beta = 0.071\), \(SE = 0.006\), likelihood-ratio \(\chi^2(1) = 158.58\), \(p < .001\)). Substantively, a 1 SD increase in corpus-average similarity, equivalent to roughly 6.5 percentage points more overlap, corresponds to about 33\% more likes than expected, while the corresponding 1 SD increase in account-average similarity, about 7.9 points more overlap, corresponds to about 18\% more likes than expected. Quadratic terms are not supported in either model (\(p = .436\) and \(p = .203\)), so the fitted relationships are adequately summarized as linear.

This suggests a slight legibility advantage: videos closer to common corpus- or account-level design patterns tend to perform better, but only because some specific kinds of similarity are rewarded while others are not. Videos closer to the wider Estonian brand TikTok repertoire tend to outperform more unusual configurations (while controlling for baseline visibility). The within-account pattern is weaker but points in the same direction: agreement with an account's own modal profile can help, but it matters less than agreement with the broader field.

The lower panels show that the aggregate slope is not an effect of uniform conformity. On the corpus side (A), positive modal-agreement effects include original spoken audio, absence of visual filters, person-focused content, and humorous tone. Several corpus defaults are more favorable when broken: agreement with the corpus-modal promotional intent predicts fewer likes than divergence from it, as do agreement with a consistent mood arc, absence of irony, and absence of stylized makeup. The account-level decomposition is weaker but similar, with positive shifts around visual-filter state, audio source, social distance, and setting, and negative shifts around mood arc, music origin, diegetic sound effects, and irony. Panels A and B should therefore be read as net summaries over heterogeneous variable-level contributions, not as evidence that every kind of similarity is equally advantageous.

Appendix Figure~\ref{fig:matchcommentsappendix} applies the same directional overlap-decomposition to comment count. As expected from the smaller count scale, the raw deltas are much smaller in absolute terms, but stable positive and negative contrasts still concentrate around directness, audio source, Labovian hook or CTA variables, and selected semiotic features.
\FloatBarrier

\subsection{Recurring Content-Side Predictors}

Table~\ref{tab:recurrentpredictors} condenses the union of the top-five CatBoost predictors across the four engagement outcomes, and reports the strongest stage-2 coefficient within each recurrent predictor variable. Because these are companion-model coefficients rather than CatBoost importance scores, they can be read as directional summaries of recurrent predictor variables, not as substitutes for predictive-contribution rankings.

\begin{table}[htbp]
\centering
\small
\caption{Recurrent top-five predictor variables across outcomes. Rows are the union of the top five CatBoost predictors for each outcome, including the two controls. For categorical predictors, each cell reports the strongest significant stage-2 effect-coded level contrast within that variable, with the coefficient shown first and the level label second. These coefficients are deviations on the \(\log(1+y)\) scale from the adjusted mean of that categorical variable in the outcome-specific mixed model, conditional on the other selected predictors and the account random intercept. Thus a value such as \(+0.10\) indicates a modest upward deviation for that level, whereas \(+0.55\) indicates a substantially larger upward deviation. `NC' indicates that the predictor variable recurred in stage 1, but no single category level emerged as a clear dominant contrast in the smaller stage-2 model for that outcome. Asterisks indicate \(^{*}p<.05\), \(^{**}p<.01\), \(^{***}p<.001\).}
\label{tab:recurrentpredictors}
\input{tables/table_recurrent_predictor_coefficients.tex}
\end{table}

At this level, audio source type is the only content variable that recurs across all four engagement outcomes. Communicative intent and setting recur most consistently thereafter; coda/CTA type and Burke's dominant element are mainly comment-specific, while primary topic is more tied to shares and saves. At the broader top-ten level, video format, audience address, and Labovian hook variables also recur. The strongest predictors are therefore not deep story templates, but mid-level packaging variables: how the video addresses the viewer, organizes sound, establishes setting, and manages its opening or closing frame.

Table~\ref{tab:recurrentpredictors} is a joint-model summary, so each variable is evaluated in the presence of other selected predictors. Stronger or redundant variables can absorb signal that would appear larger in isolation; Appendix Table~\ref{tab:univariatescreen} gives the complementary one-variable-at-a-time view (with control for visibility). The directional coefficients nevertheless show clear outcome specialization. Likes are associated with direct viewer orientation and legible settings; unclear hooks, abstract or interview-style formats, and LLM-coded trending or commercial music are negatively associated with likes. Comments are the most dialogic outcome: engagement-bait codas dominate, and purposive Burkean framing is positive, although the latter should be read cautiously because the variable falls into the lower validation-reliability tier. Shares favor original spoken audio and situated presentation more than generic trend participation, while saves favor direct address, niche-directed hooks, and recognizable framing.

Across outcomes, the variables that transfer most consistently from interpretive frameworks into predictive short-video models describe audience relation, audiovisual organization, and the narrative management of attention. The more archetypal Propp/Campbell and Polti abstractions do not recur among the strongest predictors. Supplementary format-level error analysis points in the same direction: some platform genres are more regular than others, with abstract/visual-montage relatively predictable for comments and shares and performance/dance harder to predict, though these differences remain descriptive because formats differ in frequency and engagement dispersion.

\FloatBarrier

\section{Discussion}

Methodologically, the study shows that theory-grounded LLM annotation can support interpretable multimodal analysis at a scale that would be difficult to reach with manual close coding alone. Recent multimodal AI-powered studies on ads, mental-health discourse, and influencer content show that short-form video can now be processed at large scale \parencite{yang2026mllmvadstory,zhang2026decoding,zha2026interpreting,starovolovsky-shitrit2025value}. The remaining bottleneck is access to interpretable descriptions of how videos make meaning and how audiences relate to them. Rather than inventing these descriptors from scratch, the workflow draws on existing long traditions of scholarship in narrative, rhetoric, semiotics, and multimodal analysis, translating established interpretive frameworks into structured variables that can be inspected, challenged, revised, and modeled. This supports both cultural-analytic mapping of patterned variation and practical diagnostics of which communicative features carry signal in a specific platform niche. It makes strategy-oriented modeling feasible without the cost of manually coding thousands of videos across dozens of variables. 
A machine-assisted quantitizing approach allows many theoretical lenses to be brought to the data at once and evaluated empirically rather than assumed in advance, while the annotation procedure and resulting variable values remain inspectable. This makes the interpretive layer more auditable than approaches that move directly from videos to holistic latent representations.

We sought to test whether relatively durable interpretive dimensions, developed well before TikTok and not tied to any single weekly or monthly trend cycle, still organize meaning and engagement in platform-native short video. 
The variables add a modest but consistent increment in held-out predictive performance on engagement, as baseline exposure from larger follower bases and time on the platform are already strong controls. The small gains in descriptive power indicate that narrative and structure play a modest but detectable role in TikTok engagement, as the models capture engagement-relevant variation not reducible to account size.

The outcome-specific models show why engagement should not be collapsed into a single virality measure: comment counts are most tied to response-soliciting codas and text overlays, likes to audio source, communicative intent, and setting, saves to direct address and clear hooks, and shares to circulation-oriented features such as original spoken audio, topic, and format.
Other variables partition the corpus in interesting and interpretable ways without strongly predicting engagement. This pattern reflects recent work emphasizing TikTok's multimodal complexity and its culture of rapid, highly legible forms \parencite{Lin2023,Grzenkowicz2025,Christiansen2025}, and aligns with engagement studies showing that second-person address, sentiment, audiovisual style, and richer multimodal feature sets matter for short-video engagement \parencite{cheng2024like,zhang2025audiovisual,xiao2025multimodal,sun2025engagement}. In the present corpus, engagement patterns are associated less with deep mythic structure than with how a clip frames its viewer relation, organizes its soundtrack, establishes a recognizable format, and manages the first and last seconds.

Several limitations remain. The sample is purposive and restricted to Estonian brand and organizational accounts; different dynamics may drive engagement in different niches of TikTok and social video platforms more broadly. The main contribution is methodological rather than variable-specific: the workflow yields interpretable, modestly predictive measures that can be extended in future work. We focused specifically on arguably more durable narrative and communicative structure rather than surface or more transient features like presenter appearance, topics, or current trends. Videos may become popular and gain engagement by doing what is already popular, riding an already visible sound, template, or meme, or conversely, by being unique or a trend-setter. Both dynamics have been observed in other domains of cumulative human culture, including cultural markets and popular culture, online diffusion, art and artistic careers, and cultural evolution of language \parencite{salganik_experimental_2006,bentley_regular_2007,weng_virality_2013,goel_structural_2016,fraiberger_quantifying_2018,karjus_compression_2023,liu_hot_2018,karjus_quantifying_2020}.

The results also depend on one prompt, one LLM, and one operationalization of the underlying theories. Recent methodological critiques warn about validity, reliability, replicability, and social or multilingual bias in LLM annotation pipelines \parencite{dunivin2025scaling,lin2025navigating,cui2025bias}. We did not focus on model comparison or competing implementations of the theories in this proof-of-concept contribution, but both are relevant in basic-science settings, where variation across operationalizations is itself an object of study, and in applied settings, where model choice requires balancing accuracy, cost, and deployment constraints.

The reported human validation is exploratory rather than definitive, as it relied on a small sample of videos and non-specialist annotators with modest inter-rater agreement. The reliability gradient it suggests should therefore be treated as a working caution rather than a final benchmark. A larger validation with domain-expert annotators is required before strong claims about LLM faithfulness to the underlying theoretical constructs can be made. These limitations argue for auditable workflows that preserve prompts, retain structured outputs, validate manual subsets, compare models where possible, and interpret downstream associations as model-assisted measurements rather than direct observations of latent social reality \parencite{Tai2024,Ziems2024,Karjus2025,Thapa2025}.

Future work could extend the analysis in several directions. The annotation layer should be manually validated on a targeted subset of videos and variables. The same workflow could be tested across additional sectors, languages, and platform niches to determine which predictor variables are robust and which are domain-specific. The comment material, which we explored only to a limited extent, could be incorporated more directly and annotated using deeper variables beyond sentiment, allowing comment-side response patterns to be analyzed beyond engagement counts alone.

\section{Conclusion}

This study shows that theory-grounded LLM annotation can turn interpretive concepts from narrative, rhetoric, semiotics, and multimodal analysis into auditable variables for short-form video research. In the Estonian brand and organizational TikTok corpus, the approach reveals a structured platform repertoire organized around promotion, entertainment, explanation, and recognizable modes of audiovisual presentation. The same variables add moderate but consistent predictive value for video likes, comments, shares, and saves beyond follower count and video age controls, with the clearest signals coming from broad communicative dimensions of how videos address viewers, establish purpose and format, and manage their opening and closing moments. Engagement also proves multidimensional, with attention, conversation, circulation, and saving linked to partly different aspects of video design. Content configurations closer to familiar corpus- and account-level repertoires tend to perform somewhat better in our sample, although this advantage comes from specific kinds of legibility rather than blanket conformity. The workflow's value lies in the combination of scale and interpretability: it can screen a broad theory-derived codebook, identify which dimensions carry signal in a given platform niche, and support both cultural-analytic comparison and creator-facing guidance.

\FloatBarrier

\section*{Declarations}

\subsection*{Author contributions}
S.P. collected the data, carried out the video annotation process, and wrote parts of the paper. A.K. analyzed the data, carried out further annotation, designed the graphs, and wrote the paper. We further acknowledge the help of Ellu-Marie Meos and Martin Karjus in constructing the list of accounts and initial project ideation.

\subsection*{Funding}
This work was supported by the Tallinn University School of Humanities research support grant "Sotsiaalmeedia suurandmete analüüs tehisintellekti tööriistadega". A.~Karjus was further co-funded by the European Union through the European Regional Development Fund under the Sectoral Mobility measure (SekMo), project number 2021-2027.1.01.25-1349.

\subsection*{Data and code availability}

The code and scraping instructions, as well as annotation prompt, schema, and derived analysis outputs are available at
\url{https://github.com/SanderPaekivi/tiktok_scraper_gemini_annotator}.
The original videos and comment text are not redistributed here.

\printbibliography
\FloatBarrier
\newpage

\section*{Supplementary Appendix}

\renewcommand{\thefigure}{S\arabic{figure}}
\setcounter{figure}{0}
\renewcommand{\thesection}{S\arabic{section} }  
\setcounter{section}{0}

\section{Exploratory Co-occurrence Matrices}

\begin{figure}[htbp]
\centering
\includegraphics[width=\textwidth]{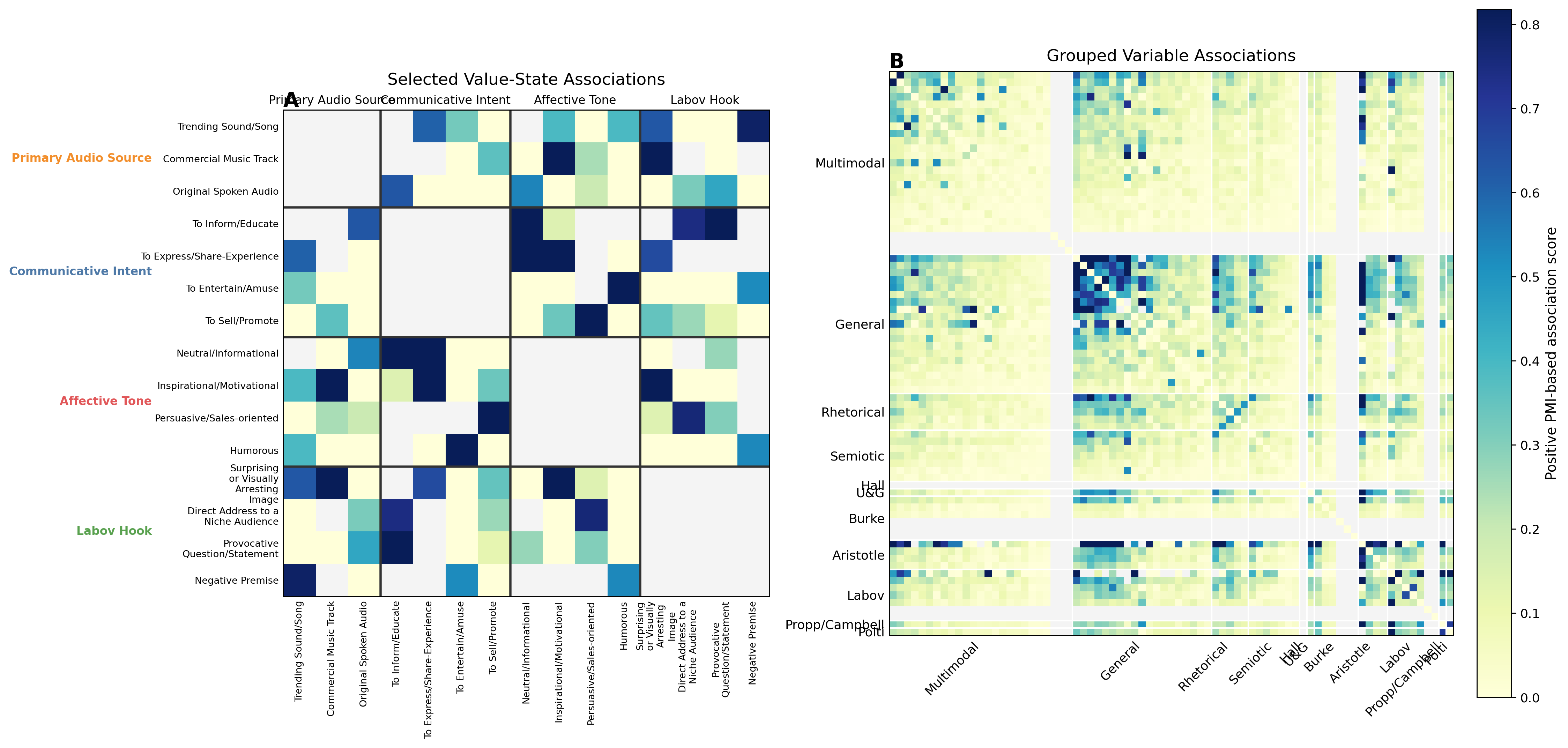}
\caption{Supplementary co-occurrence structure of the annotation space. Panel A shows selected value-state associations for four interpretable variables, with levels grouped by predictor variable and colored by theory block. Panel B aggregates the same distinctiveness logic to the grouped-variable level, with theory blocks separated along both axes. In both panels, darker cells indicate stronger over-represented pairings on a weighted positive PMI scale relative to marginal frequencies. Blank blocks in Panel A mark mutually exclusive levels from the same predictor variable. Jaccard overlap is computed in the edge tables but is not used in the plotted matrices.}
\label{fig:cooccurrence_appendix}
\end{figure}

Figure \ref{fig:cooccurrence_appendix} gives a more explicit view of how codebook variables co-occur at the level of value states and grouped variables. 
Collapsed value states are linked when they co-occur in the same video, producing an edge table with raw counts, lift, pointwise mutual information (PMI), and Jaccard overlap. These value-state associations are summarized in two matrix views. Panel A retains concrete value states for a selected set of interpretable variables, ordered by predictor variable, and each cell reports the weighted mean positive PMI for one value-state pairing. Panel B aggregates the same value-state logic to the grouped-variable level, so each cell reports the weighted mean positive PMI across substantive value-state pairs linking two variables. Both panels share the same color scale, while Jaccard is retained only as a supplementary overlap statistic in the output tables. These descriptive analyses provide a compact view of how the codebook behaves as an ensemble and how the videos populate the resulting configuration space.

In the selected value-state panel, distinctive pairings include informative intent with neutral or informational tone, humorous or inspirational tone with their matching communicative framings, and specific Labovian hook strategies with corresponding soundtrack or tone choices. At the grouped-variable level, the strongest cells again cluster around video format and related presentation variables, indicating that recognizable TikTok genres come bundled with characteristic topics, delivery modes, and affective framings. These matrices complement the predictive results by showing that the annotation system recovers structured multimodal packages rather than isolated labels.

\section{Directional Overlap Effects for Comment Count}

\begin{figure}[htbp]
\centering
\includegraphics[width=\textwidth]{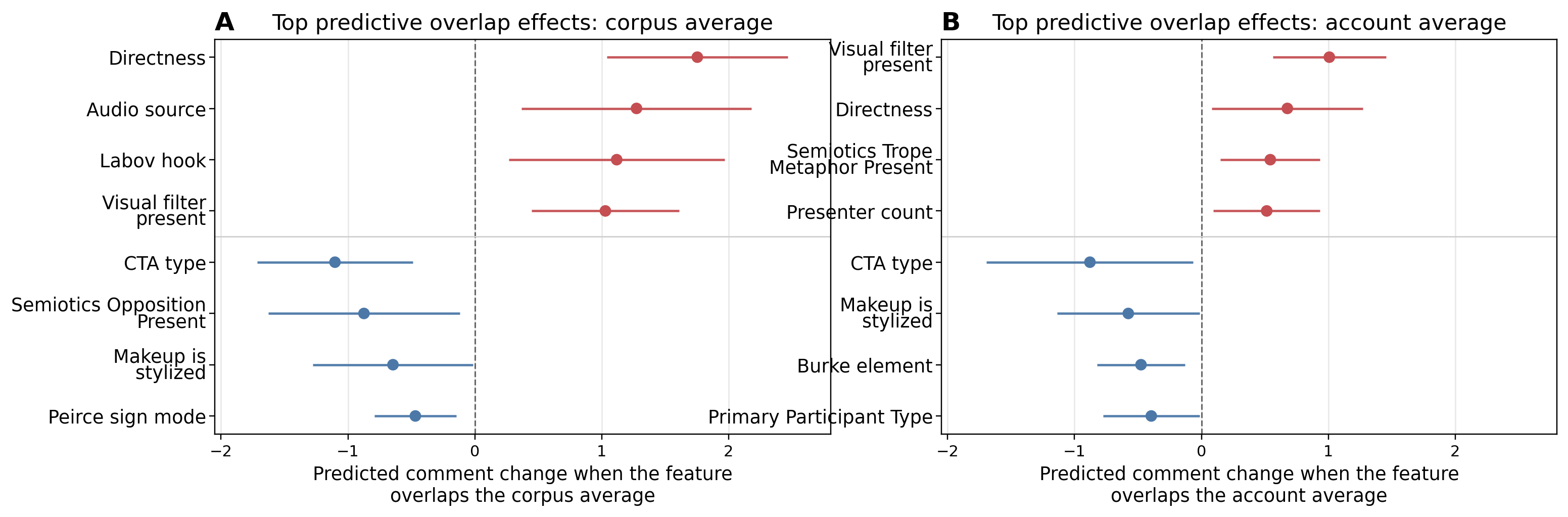}
\caption{Directional overlap effects for comment count. Using the same grouped account-split CatBoost scenario logic as Figure~\ref{fig:distinctivenesslikes}C--D, the panels show the four strongest positive and negative stable variables for corpus-average and account-average overlap when the outcome is comment count rather than likes. Points show the average predicted change in comments when a given overlap indicator is toggled from 0 to 1 on held-out videos, and horizontal lines show approximate split-based 95\% intervals. Red points indicate that overlap with the average level raises predicted comment count, while blue points indicate that divergence is more favorable. Because the raw comment-count scale is small and zero-heavy, these values should be read as modest but repeatable marginal shifts rather than as large absolute gains.}
\label{fig:matchcommentsappendix}
\end{figure}

The comment-count decomposition yields a pattern that is directionally consistent with the main-text comment analyses, but on a much smaller raw scale. Matching direct address, original spoken audio, provocative hook structures, and the absence of visual filters is associated with modest comment gains, while matching the default absence of a direct CTA and several semiotic or participant-configuration defaults is associated with lower predicted comment counts. These appendix panels are therefore best read as a variable-level diagnostic of the broader comment-count patterns rather than as a stand-alone behavioral model.

\section{One-Variable Comparisons}

Table~\ref{tab:univariatescreen} gives a one-variable-at-a-time view of how far engagement outcomes move across the common levels of individual coded variables after follower count, video age, and account differences are taken into account. The purpose is descriptive: it shows the magnitude of outcome variation associated with each design dimension on its own once those baseline differences are held constant.

For likes, comments, shares, and saves, each row comes from a pair of linear mixed-effects models on the \(\log(1+y)\) outcome scale: a baseline model with follower count and video age, and a full model that adds one coded variable, with an account random intercept in both cases. The final column uses the existing supportive-versus-other comment-activity coding and estimates the same one-variable comparison with a binomial model on the comment-coded subset, again controlling for follower count and video age while accounting for account differences. The table value is then derived from adjusted predictions on the original outcome scale.

For each variable, the highest-performing common level is identified and compared with the mean prediction across the remaining common levels. This contrast is intermediate between two simpler summaries: a best-versus-observed-mix contrast depends strongly on how common the best level already is, while a pure min-max spread can be driven by a single worst level. The present contrast retains a best-common-level counterfactual interpretation while avoiding both of those distortions. Each cell reports that best-versus-rest contrast as a percentage of the model-implied average outcome under the observed mix of that variable's levels. The same percentage interpretation carries across comments, shares, saves, and the supportive-comment column, which keeps the columns on a comparable scale. Variables enter the table only when at least two levels each occur in at least 100 videos and at least five accounts, which excludes effectively one-sided predictors; the few blank cells mark cases in which the restricted supportive-comment model still did not yield a stable estimate. Stars mark Bonferroni-corrected likelihood-ratio comparisons between the baseline and full models across the full set of variable-outcome tests reported in the table, so they should be read as a conservative reliability cue rather than as the main substantive ranking. Theory blocks are ordered by their average likes contrast, and variables within each block are ordered by the likes column.

\refstepcounter{table}\label{tab:univariatescreen}
\noindent\textbf{Table \thetable.} One-variable-at-a-time adjusted contrasts by variable, expressed as percentages of the adjusted outcome mean. Each cell shows the contrast between the highest-performing common level of that variable and the mean of its remaining common levels after adjustment for follower count, video age, and account differences; for example, the first likes value, 30, means a contrast equal to about 30\% of the average adjusted likes level. Stars mark Bonferroni-corrected likelihood-ratio comparisons between the baseline and full models across the full table of variable-outcome tests.

\input{tables/table_univariate_screen.tex}

\end{document}

%% file: tables/table_recurrent_predictor_coefficients.tex
{\renewcommand{\arraystretch}{1.1}
\begin{tabularx}{\textwidth}{>{\raggedright\arraybackslash}p{0.22\textwidth} >{\raggedright\arraybackslash}X >{\raggedright\arraybackslash}X >{\raggedright\arraybackslash}X >{\raggedright\arraybackslash}X}
\toprule
Predictor variable & Likes & Comments & Shares & Saves \\
\midrule
Coda / CTA type & NC & +1.00*** Engagement Bait & NC & +0.21*** Engagement Bait \\
Audio source type & -0.38** Trending Sound/Song & NC & +0.45* Original Spoken Audio & -0.55*** Trending Sound/Song \\
Primary topic & NC & NC & -0.40* Business/Finance & -0.48** Politics/Social-Commentary \\
Setting location & +0.34*** Outdoors-Urban & +0.19* Outdoors-Urban & +0.36*** Outdoors-Urban & +0.27** Outdoors-Urban \\
Communicative intent & +0.17** To Entertain/Amuse & NC & -0.28* To Persuade/Convince & NC \\
Burke dominant element & NC & +0.22*** Purpose & NC & NC \\
Log follower count & +0.50*** & -0.01 & -0.03 & +0.07 \\
Log video age (days) & +0.07*** & +0.01 & -0.20*** & +0.04** \\
\bottomrule
\end{tabularx}
}

%% file: tables/table_univariate_screen.tex
\setlength{\LTleft}{0pt}
\setlength{\LTright}{0pt}
\footnotesize
\begin{longtable}{>{\raggedright\setlength{\parindent}{0pt}\hangindent=1.2em\hangafter=1\arraybackslash}m{0.34\textwidth} >{\raggedright\arraybackslash}m{0.09\textwidth} >{\raggedright\arraybackslash}m{0.09\textwidth} >{\raggedright\arraybackslash}m{0.09\textwidth} >{\raggedright\arraybackslash}m{0.09\textwidth} >{\raggedright\arraybackslash}m{0.15\textwidth}}
\toprule
Variable & Likes & Comments & Shares & Saves & Supportive comments \\
\midrule
\endfirsthead
\toprule
Variable & Likes & Comments & Shares & Saves & Supportive comments \\
\midrule
\endhead
\midrule
\multicolumn{6}{r}{\emph{Continued on next page}} \\
\endfoot
\bottomrule
\endlastfoot
\midrule[1.1pt]
\multicolumn{1}{l}{\textbf{Propp/Campbell}} & & & & & \\
Narrative Arc Shape & $30^{***}$ & $34^{***}$ & $35^{***}$ & $39^{***}$ & -- \\
\midrule[1.1pt]
\multicolumn{1}{l}{\textbf{Rhetorical}} & & & & & \\
Directness & $38^{***}$ & $61^{***}$ & $38^{***}$ & $45^{***}$ & $18^{***}$ \\
Tone & $37^{***}$ & $34^{***}$ & $44^{***}$ & $18^{***}$ & -- \\
Address Social Distance & $28^{***}$ & $17$ & $32^{***}$ & $21^{**}$ & $19$ \\
Direct Gaze Present & $24^{***}$ & $36^{***}$ & $18^{**}$ & $23^{***}$ & $5$ \\
Address power dynamic & $20^{***}$ & $14$ & $20^{***}$ & $25$ & $29^{***}$ \\
\midrule[1.1pt]
\multicolumn{1}{l}{\textbf{General}} & & & & & \\
Intent & $53^{***}$ & $17$ & $62^{***}$ & $40^{***}$ & $31^{***}$ \\
Makeup is stylized & $50^{***}$ & $45^{***}$ & $56^{***}$ & $54^{***}$ & $21^{**}$ \\
Presentation & $47^{***}$ & $29^{***}$ & $33^{***}$ & $39^{***}$ & $16^{***}$ \\
Content focus & $46^{***}$ & $33^{***}$ & $40^{***}$ & $34^{***}$ & $2$ \\
Topic & $45^{***}$ & $67^{***}$ & $78^{***}$ & $34^{***}$ & $22^{***}$ \\
Overall mood & $42^{***}$ & $29^{***}$ & $60^{***}$ & $37^{***}$ & $36^{***}$ \\
Setting & $29^{***}$ & $20^{***}$ & $41^{***}$ & $27^{***}$ & $12$ \\
Delivery & $28^{***}$ & $39^{***}$ & $83^{***}$ & $24^{***}$ & $22^{***}$ \\
Presenter Primary Activity & $27^{***}$ & $48^{***}$ & $33^{***}$ & $34^{***}$ & -- \\
Reference Material Present & $26^{*}$ & $29$ & $13$ & $26$ & $12$ \\
Genre & $25^{***}$ & $67^{***}$ & $50^{***}$ & $51^{***}$ & $40^{***}$ \\
Presenter count & $23^{***}$ & $27^{***}$ & $23^{***}$ & $17^{***}$ & $11^{**}$ \\
Presenter Is Costumed & $22$ & $18$ & $35^{*}$ & $18$ & $17^{*}$ \\
\shortstack[l]{Presenter Is Digitally\\\hspace*{0.8em}Altered} & $20$ & $18$ & $22$ & $18$ & $0$ \\
Presenter demographic & $17^{***}$ & $38^{***}$ & $27^{***}$ & $12^{***}$ & $8$ \\
Presenter Attire Style & $16^{***}$ & $7^{**}$ & $26^{***}$ & $12^{***}$ & $9$ \\
Mood Arc & $13^{***}$ & $28$ & $28^{***}$ & $30^{***}$ & $13$ \\
Perceived Authenticity & $9$ & $25^{***}$ & $26^{***}$ & $12$ & $17$ \\
Setting is specific & $8$ & $13$ & $10$ & $10$ & $6$ \\
\midrule[1.1pt]
\multicolumn{1}{l}{\textbf{Multimodal}} & & & & & \\
Audio source & $55^{***}$ & $49^{***}$ & $64^{***}$ & $66^{***}$ & $18^{***}$ \\
Visual filter present & $40^{***}$ & $39^{***}$ & $44^{***}$ & $36^{***}$ & $1$ \\
Text overlay function & $39^{***}$ & $168^{***}$ & $41^{***}$ & $40^{***}$ & $36^{***}$ \\
Audio layer & $39^{***}$ & $40^{***}$ & $51^{***}$ & $53^{***}$ & $13$ \\
Camera Angle & $35^{***}$ & $27^{***}$ & $28^{*}$ & $32^{***}$ & $12$ \\
Music Origin & $34^{***}$ & $11$ & $15^{***}$ & $34^{***}$ & $37^{**}$ \\
Editing Pace & $34^{***}$ & $31^{***}$ & $41^{***}$ & $39^{***}$ & $1$ \\
Participant role & $32^{***}$ & $22$ & $32^{***}$ & $15$ & $3$ \\
Music Emotion Cowen & $29^{***}$ & $27^{***}$ & $40^{***}$ & $55^{***}$ & -- \\
Framing & $28^{***}$ & $18^{***}$ & $21^{***}$ & $25^{***}$ & $7$ \\
Primary Participant Type & $28^{***}$ & $27^{***}$ & $28^{***}$ & $18^{***}$ & $7$ \\
Non Diegetic Sfx Present & $24^{***}$ & $15$ & $21$ & $23^{***}$ & $5$ \\
Dominant Camera Movement & $24^{***}$ & $27^{**}$ & $20$ & $21^{***}$ & -- \\
Speed Alteration Present & $22$ & $18$ & $27$ & $26$ & -- \\
Av Sync Present & $19^{***}$ & $15^{**}$ & $24^{***}$ & $26^{***}$ & $10$ \\
Diegetic Sfx Present & $16^{**}$ & $5$ & $18$ & $21^{***}$ & $2$ \\
Composition & $16$ & $17$ & $15$ & $19^{*}$ & $7$ \\
Dominant Transition Type & $14^{***}$ & $7$ & $17$ & $29^{***}$ & $1$ \\
Jump Cuts Prominent & $6$ & $11$ & $3$ & $2$ & $6$ \\
Is Dynamic Editing & $5$ & $1$ & $2$ & $14^{**}$ & $2$ \\
\shortstack[l]{Extradiegetic Graphics\\\hspace*{0.8em}Present} & $2$ & $9$ & $5$ & $12^{*}$ & $4$ \\
\midrule[1.1pt]
\multicolumn{1}{l}{\textbf{Labov}} & & & & & \\
\shortstack[l]{Labov Complicating Action\\\hspace*{0.8em}Present} & $45^{***}$ & $31^{*}$ & $42^{***}$ & $47^{***}$ & $1$ \\
CTA type & $42^{***}$ & $285^{***}$ & $21^{***}$ & $60^{***}$ & $19^{**}$ \\
Labov hook & $20^{***}$ & $37^{***}$ & $25^{***}$ & $32^{***}$ & $14^{***}$ \\
Labov Orientation Present & $19$ & $10$ & $29$ & $27$ & -- \\
Labov Resolution Present & $15^{***}$ & $5$ & $13$ & $19^{***}$ & $5$ \\
Labov Coda Present & $3$ & $24^{***}$ & $5$ & $15^{***}$ & $3$ \\
\midrule[1.1pt]
\multicolumn{1}{l}{\textbf{Semiotic}} & & & & & \\
\shortstack[l]{Semiotics Trope Irony\\\hspace*{0.8em}Present} & $42^{***}$ & $9$ & $58^{***}$ & $25^{***}$ & $4$ \\
\shortstack[l]{Semiotics Trope Metonymy\\\hspace*{0.8em}Present} & $27^{***}$ & $18^{**}$ & $39^{***}$ & $37^{***}$ & $4$ \\
Semiotics Anchorage Status & $22$ & $17$ & $36^{***}$ & $42^{***}$ & $12$ \\
Peirce sign mode & $19^{***}$ & $36^{***}$ & $35^{***}$ & $25^{***}$ & $2$ \\
\shortstack[l]{Semiotics Opposition\\\hspace*{0.8em}Present} & $19^{***}$ & $13$ & $25^{***}$ & $9$ & $12^{*}$ \\
\shortstack[l]{Semiotics Trope Synecdoche\\\hspace*{0.8em}Present} & $18^{***}$ & $19^{*}$ & $19$ & $17^{*}$ & $4$ \\
\shortstack[l]{Semiotics Trope Metaphor\\\hspace*{0.8em}Present} & $15^{***}$ & $21^{***}$ & $9$ & $17^{***}$ & $4$ \\
\midrule[1.1pt]
\multicolumn{1}{l}{\textbf{U\&G}} & & & & & \\
Gratification & $23^{***}$ & $16$ & $24^{***}$ & $28^{***}$ & $21^{***}$ \\
\midrule[1.1pt]
\multicolumn{1}{l}{\textbf{Polti}} & & & & & \\
Polti Core Conflict & $20^{***}$ & $15^{***}$ & $33^{***}$ & $15^{***}$ & $10$ \\
\midrule[1.1pt]
\multicolumn{1}{l}{\textbf{Aristotle}} & & & & & \\
Appeal Pathos Present & $24$ & $3$ & $30$ & $7$ & $48^{***}$ \\
Appeal Ethos Present & $13^{**}$ & $11$ & $20^{***}$ & $5$ & $12^{**}$ \\
Rhetorical appeal & $11$ & $20$ & $12$ & $40^{***}$ & $37^{***}$ \\
Appeal Logos Present & $3$ & $21^{***}$ & $3$ & $26^{***}$ & $29^{***}$ \\
\midrule[1.1pt]
\multicolumn{1}{l}{\textbf{Burke}} & & & & & \\
Burke element & $17^{***}$ & $64^{***}$ & $25^{***}$ & $10^{***}$ & $13^{***}$ \\
Burke Purpose Present & $11$ & $16$ & $4$ & $28^{***}$ & $10$ \\
Burke Agency Present & $3$ & $15$ & $10$ & $15$ & $1$ \\
\end{longtable}